\documentclass[fleqn,usenatbib]{mnras}

\usepackage{newtxtext,newtxmath}
\usepackage[T1]{fontenc}
\DeclareRobustCommand{\VAN}[3]{#2}
\let\VANthebibliography\thebibliography
\def\thebibliography{\DeclareRobustCommand{\VAN}[3]{##3}\VANthebibliography}

\usepackage{amsmath}
\usepackage{graphicx}
\usepackage{aliascnt}
\usepackage{enumitem}
\usepackage{xcolor}
\usepackage{hyperref}
\usepackage{gensymb}
\usepackage{CJK}
\usepackage{tabularx}
\usepackage{orcidlink}
\usepackage{fontawesome}
\usepackage[normalem]{ulem}
\hypersetup{
    unicode, 
    colorlinks=true,
    linkcolor=linkcolor,
    citecolor=linkcolor,
    filecolor=linkcolor,
    urlcolor=linkcolor,
}
\definecolor{linkcolor}{rgb}{0.0,0.3,0.5}

\defcitealias{Buder2025c}{Paper~I}

\title[Were Splash stars heated or already born hot?]{The chemodynamical memory of a major merger in a NIHAO-UHD Milky Way analogue -- II. Were Splash stars heated or already born hot?}

\author[S. Buder et al.]{Sven Buder,$^{1}$\thanks{E-mail: sven.buder@anu.edu.au}\thanks{Australian Research Council DECRA Fellow}\orcidlink{0000-0002-4031-8553}
Tobias Buck,$^{2,3}$\orcidlink{0000-0003-2027-399X}
Ása Skúladóttir,$^{4}$\orcidlink{0000-0001-9155-9018}
Melissa Ness,$^{1}$\orcidlink{0000-0001-5082-6693}
Madeleine McKenzie,$^{5}$\thanks{NASA Hubble Fellow}\orcidlink{0000-0002-1715-1257}
and\newauthor
Stephanie Monty$^{6, 7}$\orcidlink{0000-0002-9225-5822}
\\
$^{1}$Research School of Astronomy and Astrophysics, Australian National University, Canberra, ACT 2611, Australia\\
$^{2}$Universit{\"a}t Heidelberg, Interdisziplin{\"a}res Zentrum f{\"u}r Wissenschaftliches Rechnen, Im Neuenheimer Feld 205, D-69120 Heidelberg, Germany\\
$^{3}$Universit{\"a}t Heidelberg, Zentrum f{\"u}r Astronomie, Institut f{\"u}r Theoretische Astrophysik, Albert-Ueberle-Straße 2, D-69120 Heidelberg, Germany\\
$^{4}$Dipartimento di Fisica e Astronomia, Universitá degli Studi di Firenze, Via G. Sansone 1, I-50019 Sesto Fiorentino, Italy\\
$^{5}$The Observatories of the Carnegie Institution for Science, 813 Santa Barbara Street, Pasadena, 91101, CA, USA\\
$^{6}$Institute of Astronomy, University of Cambridge, Madingley Road, Cambridge CB3 0HA, UK\\
$^{7}$Department of Astronomy, New Mexico State University, Las Cruces, NM 88003, USA
}

\date{Accepted 2025 April 14. Received 2025 April 6; in original form 2025 October 23}

\pubyear{\the\year{}}

\begin{document}
\label{firstpage}
\pagerange{\pageref{firstpage}--\pageref{lastpage}}
\maketitle

\begin{abstract} 
One of the most debated consequences of the Milky Way's last major merger is the so-called \textit{Splash}: stars with disc-like chemistry but halo-like kinematics, often interpreted as evidence for the violent heating of an early protodisc. Using the same high-resolution NIHAO-UHD cosmological simulation analysed in Paper~I, we test whether, and if so how, a Splash-like population arises in the Milky Way analogue. By tracing stellar birth positions, ages, and present-day orbits, we find that protodisc stars were already born on dynamically hot orbits, with only limited additional dynamical \textit{splashing} of these particular in-situ stars despite a 1:5 stellar mass merger. A subset of stars, particularly those that end up in the Solar neighbourhood, shows evidence for merger-driven angular-momentum redistribution, but the overall kinematic distribution of stars with Splash-like chemistry remains largely unchanged. The observed Splash may therefore primarily reflect the already turbulent early disc, subsequently intermixed with accreted stars and those formed from merger-driven gas inflows, rather than a distinct merger-heated population. When selecting stars with similar chemistry and age as the Splash-like ones, we find their azimuthal velocity distribution to be broad and positively skewed, with $V_\varphi = 73_{-59}^{+74}\,\mathrm{km\,s^{-1}}$. The transition to a rotation-supported disc with large azimuthal velocities occurs only during or after the merger. Our results suggest an alternative to the proposed splashing scenario and highlight the need to disentangle the relative contributions of merger-induced heating and intrinsically hot disc formation to clarify the nature of Splash-like stars and their role in shaping the early Milky Way.
\end{abstract}

\begin{keywords}
Galaxy: abundances -- Galaxy: evolution -- Galaxy: formation -- Galaxy: kinematics and dynamics -- Galaxy: structure
\end{keywords}


\section{Introduction}
\label{sec:introduction}

A long-standing goal of Galactic archaeology is to reconstruct how mergers shaped the early Milky Way. Thanks to \textit{Gaia} astrometry \citep{Brown2021b} and large spectroscopic surveys \citep{Jofre2019}, it is now clear that our Galaxy experienced at least one major accretion event about 8--10~Gyr ago, commonly referred to as the Gaia-Sausage-Enceladus (hereafter \textit{GSE}) merger \citep{Haywood2018b, Belokurov2018, Helmi2018, Naidu2020}. This event left an imprint in the stellar halo, with estimates suggesting that the \textit{GSE} progenitor contributes approximately 20–65\% of halo stars depending on location within the halo \citep{Naidu2020}. Growing evidence suggests that it also influenced a pre-existing disc, dynamically mixing the earliest in-situ populations with accreted debris \citep[see review by][]{Helmi2020}.

One of the most debated signatures of this process is the so-called Splash population \citep{Belokurov2020, Belokurov2022}. These stars appear chemically similar to the old, $\upalpha$-enhanced thick disc, but occupy hotter, halo-like orbits. Their dual nature has made them central to arguments about whether the \textit{GSE} merger violently heated the protodisc, or whether some old stars were already born on dynamically hot orbits in our Galaxy’s turbulent youth.

Observational studies first pointed to an in-situ halo component overlapping with the disc in chemistry. \citet{Bonaca2017} identified metal-rich, prograde halo stars consistent with a heated thick disc origin. \citet{Gallart2019} analysed the ages of this old in-situ population with halo-like kinematics, arguing that its stars are among the oldest in the Milky Way and likely predate much of the thick disc. Expanding on this, \citet{Bonaca2020} used \textit{Gaia} DR2 and H3 survey data to argue that disc-like chemistry but halo-like kinematics reflect merger-driven heating, providing a timeline for disc perturbation shortly after \textit{GSE}. Similarly, \citet{DiMatteo2019,DiMatteo2020} interpreted metal-rich, eccentric stars as evidence that the \textit{GSE} merger dynamically heated pre-existing disc stars into the halo, strengthening the case that Splash stars trace this violent epoch.

Other studies have argued for alternative or complementary origins. In a hydrodynamical simulation of an isolated clumpy disc, \citet{Amarante2020} showed that a Splash-like component can form without mergers: stars scattered by giant clumps during early, turbulent disc phases naturally occupy low-angular-momentum, halo-like orbits, yet remain chemically continuous with the thick disc. More recently, cosmological simulations have broadened the picture. \citet{Grand2020} showed that a gas-rich GSE-like merger can both heat a pre-existing proto-disc and trigger a compact starburst component, producing Splash-like stars with chemistry overlapping the thick disc and naturally generating chemo-kinematic trends seen in the Milky Way. Similar gas-rich merger scenarios have also been invoked to explain early chemical enrichment patterns in the Galactic disc \citep{Ciuca2024}. \citet{Dillamore2022} found that Splash-like stars could be a generic feature of Milky Way analogues, while \citet{Dillamore2023} showed that bar resonances can also create halo-like substructures. Extending this, \citet{Dillamore2025} demonstrated that secular evolution can reshape integral-of-motion space, complicating attempts to uniquely attribute Splash stars to merger heating. Observationally, \citet{Kisku2025} showed that Splash stars occupy the high-$\alpha$, high-[Al,K/Fe], low-[Mn/Fe] extension of the thick disc sequence, and, in combination with ARTEMIS simulations \citep{Font2020}, argued that both major and minor mergers can produce Splash-like signatures, in agreement with the conclusions from the HESTIA simulation by \citet{Khoperskov2023}. A complementary, physically explicit pathway is feedback-driven \textit{baryon sloshing} in gas-rich discs, which isotropically heats newly formed stars and builds a thick disc without requiring a contemporaneous major heating event \citep{BlandHawthorn2025}.

Taken together, these studies highlight that Splash stars remain a crucial but unresolved tracer of the Milky Way’s formative history. They may represent violently heated disc stars, stars born hot in a turbulent proto-disc, or a mixture of both, possibly reshaped by subsequent secular evolution. The distinction matters: if Splash stars are heated, they provide a dynamical clock for the last major merger; if they are born hot, they instead reflect the conditions of the early disc, independent of mergers.

In \citet[][hereafter Paper~I]{Buder2025c}, we used a NIHAO-UHD\footnote{NIHAO-UHD is the Ultra High Definition re-run of a Milky Way analogue from the \textit{Numerical Investigation of a Hundred Astronomical Objects} (NIHAO) suite \citep{Wang2015}.} cosmological simulation of a Milky Way analogue \citep{Buck2020, Buck2021} to examine the chemodynamical memory of a major merger. \citetalias{Buder2025c} focused on the efficiency of recovering accreted stars in integral-of-motion space, the preservation of their birth radii, and chemical evolution. Building on this framework, this work (Paper~II) asks whether a Splash-like population emerges in a fully cosmological Milky Way analogue, and if so, whether these stars were \emph{heated by the merger or already born hot}. Our guiding questions are:
\begin{enumerate}[leftmargin=2em,labelwidth=0em]
    \item Does a distinct Splash population emerge in a high-resolution cosmological simulation?
    \item Is this population born in the Milky Way analogue or also/only the \textit{GSE}-like galaxy that merged with it around $8.6\,\mathrm{Gyr}$ ago?
    \item Are Splash-like stars dynamically heated, or do they form already hot in the turbulent early disc?
    \item What does the simulated Splash imply for interpreting the Milky Way’s merger-driven history?
\end{enumerate}

\section{Simulation data}
\label{sec:data}

We use the same high-resolution cosmological zoom-in simulation analysed in \citetalias{Buder2025c}, namely the NIHAO-UHD Milky Way analogue \texttt{g8.26e11} \citep{Wang2015, Buck2019b, Buck2020b, Buck2021}. The simulation was performed with the smoothed-particle hydrodynamics code \texttt{Gasoline2} \citep{Wadsley2017}, following the NIHAO feedback and enrichment framework \citep{Stinson2006, Stinson2013, Wang2015}. It adopts a $\Lambda$CDM cosmology as inferred by \citet{Planck2014} and includes metal-line cooling, star formation, and stellar feedback with chemical enrichment from asymptotic giant branch stars, core-collapse supernovae, and supernovae of type Ia \citep{Buck2021}. The NIHAO-UHD suite reaches ultra-high mass resolution; for \texttt{g8.26e11} the stellar particle mass is $7.5 \times 10^3\,\mathrm{M_{\odot}}$. This is comparable to the highest-resolution \textsc{Auriga} simulations (level~3; $\sim 6 \times 10^3\,\mathrm{M_{\odot}}$), and higher than the standard \textsc{Auriga} level~4 runs ($\sim 5 \times 10^4\,\mathrm{M_{\odot}}$; \citealt{Grand2017}) and the ARTEMIS simulations ($\approx 3.2 \times 10^4\,\mathrm{M_{\odot}}$ for $h=0.70$; \citealt{Font2020}).

This simulation was chosen because it closely resembles the Milky Way in its global properties, such as stellar mass and disc-dominated morphology, and experienced a last major merger at a lookback time of $\sim 8.6$~Gyr, broadly consistent with the timing inferred for the Milky Way's \textit{GSE} merger \citep{Helmi2018, Naidu2021}. At $z=0$, the main halo has a virial radius of $R_{200}=206\,\mathrm{kpc}$ and a total mass of $M_{200}=9.1\times10^{11}\,\mathrm{M_\odot}$, including $8.2\times10^{11}\,\mathrm{M_\odot}$ in dark matter, $6.4\times10^{10}\,\mathrm{M_\odot}$ in gas, and $2.3\times10^{10}\,\mathrm{M_\odot}$ in stars. The last major merger progenitor, a \textit{GSE}-like galaxy, had a stellar mass of $\sim 3\times10^9\,\mathrm{M_\odot}$ and a stellar mass ratio of roughly $1:5$ relative to the host at the time of infall, comparable to empirical estimates for the Milky Way's \textit{GSE} merger \citep[e.g.][]{Helmi2018, Naidu2021}. At a lookback time of $9.8\,\mathrm{Gyr}$, the last snapshot where both progenitors are cleanly identified prior to coalescence, we estimate total, gas, and dark-matter mass ratios of roughly $1:4$--$1:6$, indicating that the merger was of similar relative scale across components.
At redshift $z=0$, the stellar disc can be well described by two vertical components with scale heights of $\sim 0.4\,\mathrm{kpc}$ and $\sim 1.3\,\mathrm{kpc}$, indicating the presence of distinct thin- and thick-disc components.
A key structural property of \texttt{g8.26e11} is that it forms a multi-armed spiral disc but no significant present-day bar \citep{Buder2025}. This means that bar-driven resonances, proposed as a possible driver of the \textit{Splash} by \citet{Dillamore2022, Dillamore2023}, are likely not responsible for the features we analyse here. This is further supported by the vertical distribution of stellar populations, with younger stars ($<8\,\mathrm{Gyr}$) exhibiting a dispersion of $\sim 0.5\,\mathrm{kpc}$ and older stars ($>10\,\mathrm{Gyr}$) showing a larger dispersion of $\sim 1.2\,\mathrm{kpc}$.

A full description of the simulation setup, feedback and enrichment prescriptions, and comparison to Milky Way constraints is given in Sec.~2.1 of \citetalias{Buder2025c}. Sec.~2.2 of \citetalias{Buder2025c} details how we record stellar birth positions. In brief, we identify the first simulation snapshot in which each star particle appears and record its position relative to the evolving centre of mass of the main halo, thereby recording stellar birth locations sampled at the $\sim100\,\mathrm{Myr}$ cadence of the simulation snapshots.

Combined with the orbital energies and actions (Sec.~2.3 of \citetalias{Buder2025c}), which we computed with the \textsc{agama} code \citep{Vasiliev2019b}, this allows us to characterise the present-day orbital properties of the stars. The orbital quantities are derived by integrating stellar orbits in a multipole expansion of the simulation potential constructed from the dark matter, stellar, and gas components. Together, these quantities enable us to distinguish in-situ stars from those accreted during mergers (see Sec.~3.1 of \citetalias{Buder2025c}).

\section{Analysis: The merger process and its chemodynamical effects} \label{sec:analysis}

In \citetalias{Buder2025c}, we analysed the NIHAO-UHD Milky Way analogue to classify stars into in-situ and accreted populations, examine their orbital properties and selection efficiency in integrals-of-motion space, and evaluate how well birth positions and chemical evolution are retained as chemodynamical memory. Building on this framework, we now focus on the merger's dynamical effects. Our aim is to assess whether the simulation produces a population analogous to the observationally claimed Splash: stars born in the early disc but dynamically heated onto halo-like orbits, stars formed from merger-driven gas mixing, or stars that may be misclassified accreted populations. By linking these possibilities back to the classification, dynamical, and chemical analyses of \citetalias{Buder2025c}, we investigate whether the Splash emerges as a distinct component or overlap of several processes.

\subsection{Infall} \label{sec:analysis_infall}

In Fig.~3 of \citetalias{Buder2025c} we studied the birth positions of accreted stars within their host galaxy in timesteps of $0.1\,\mathrm{Gyr}$. The birth positions in the $X_\mathrm{birth}$ vs. $Z_\mathrm{birth}$ plane (Fig.~3b of \citetalias{Buder2025c}) indicate that the major merger galaxy followed an inclined trajectory relative to the main galaxy disc.
When tracking the orbit of the major merger galaxy, we estimate the inclination angle $\theta$ of the merger (with larger inclination meaning a more edge-on rather than face-on trajectory) to be between $34$ and $55\,\mathrm{deg}$, based on a singular value decomposition in three dimensions or a linear fit in $X_\mathrm{birth}$ vs. $Z_\mathrm{birth}$, respectively. This value is roughly comparable to the inclination of $\theta = 30\,\mathrm{deg}$ inferred by \citet{Naidu2021} for the Milky Way's \textit{GSE} merger with simulations with multiple inclination and circularity values. We note that the pattern of stars on significantly more prograde orbits (Fig.~4b of \citetalias{Buder2025c}) matches less well with the patterns of typically low absolute angular momenta observed in the Milky Way and simulated for a merger with circularity $\eta = 0.5$. Our orbit pattern rather matches the simulation outcome by \citet{Naidu2021} for a highly circular galaxy with $\eta = 0.9$ (see their Fig.~3). This relatively circular infall contrasts with the more radial trajectory typically inferred for the Milky Way's \textit{GSE} event. As a consequence, the merger is expected to induce weaker impulsive heating but stronger long-timescale torques, which may both reduce the fraction of strongly splashed stars and shift the resulting kinematic distribution towards more prograde motion. We refrain from interpreting this, however, as there are more particular merger parameters that might play an important role for energy and momentum transfer. This includes the path along which the merging galaxy spiralled into the larger galaxy relative to the galactic plane (see $X_\mathrm{birth}$ vs. $Y_\mathrm{birth}$ in Fig.~3a of \citetalias{Buder2025c}), as well as the locations of passages through this plane.
We can, however, study the kinematic/dynamical effect on already born in-situ stars as well as the chemodynamic effect on stars that formed during the merger.

\begin{figure}
    \centering
    \includegraphics[width=\columnwidth]{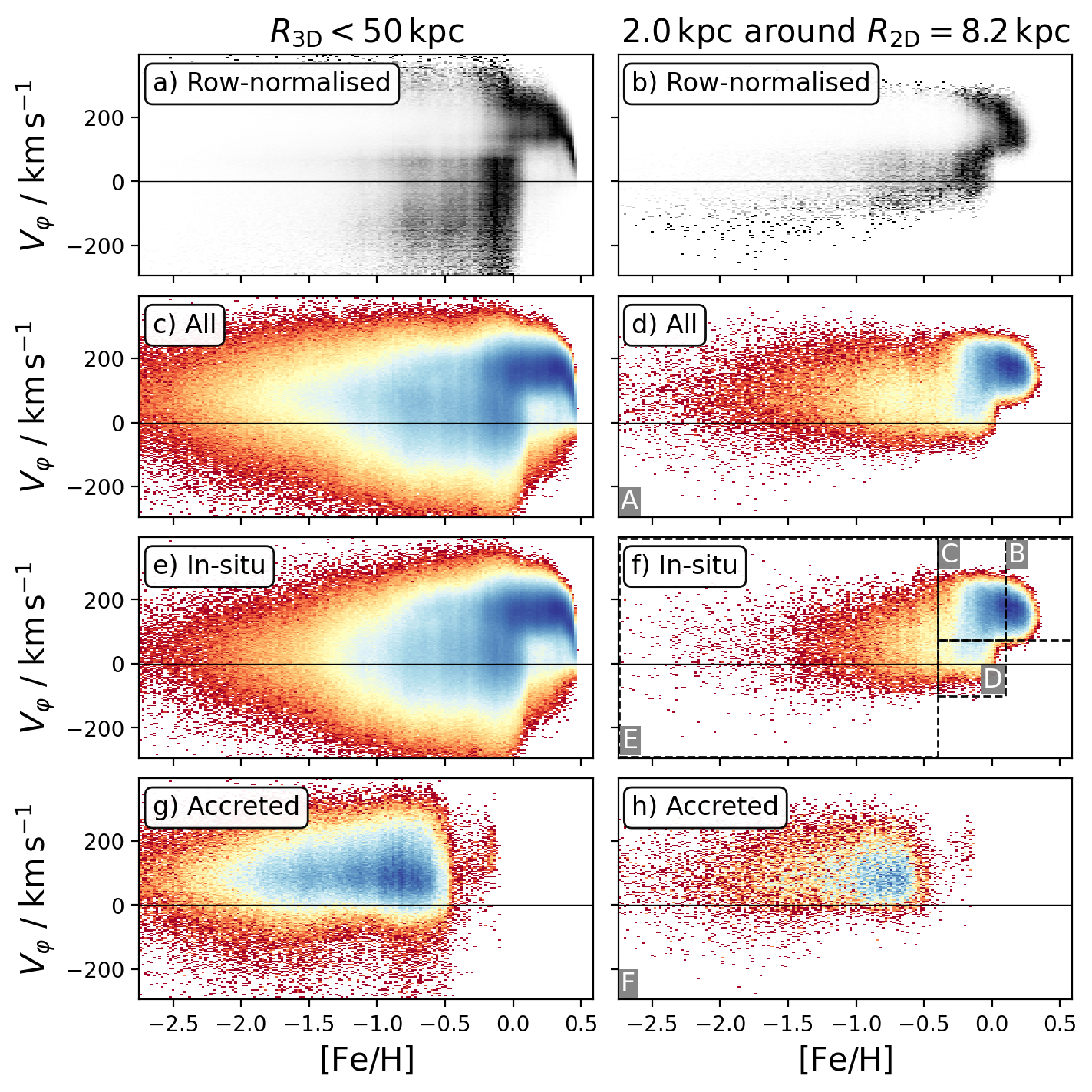}
    \caption{{Iron abundance [Fe/H]} vs. azimuthal velocity $V_\varphi$ for stars within $R_\mathrm{3D} < 50\,\mathrm{kpc}$ (left panels) and the Solar-like locus (right panels). Panels a) and b) show the density distribution, when normalising the density across each row, as done by \citet[][see their Fig~1]{Belokurov2020}. Panels c) and d) show the logarithmic density distribution of all stars. Panels e) and f) show in-situ stars, while panels g) and h) show accreted stars. Dashed black rectangles with letters A-F at their lower left corner annotate the selection of different subsamples of the Solar neighbourhood \href{https://github.com/svenbuder/gse_nihaouhd/tree/main/figures/splash_feh_vphi.png}{\faGithub}.}
    \label{fig:splash_feh_vphi}
\end{figure}

\subsection{Selection of Splash-like stars}

Based on a population of stars in the Solar neighbourhood that are more metal-rich than typical halo stars yet follow highly eccentric orbits, \citet{Belokurov2020} proposed that these stars originated in the Milky Way’s disc and were \textit{splashed} onto low-angular-momentum orbits during the last major merger. Motivated by this hypothesis, we analyse stars in our Milky Way analogue in the same plane as \citet{Belokurov2020}, namely $\mathrm{[Fe/H]}$ vs. azimuthal velocity $V_\varphi$, both for the entire galaxy and for a Solar neighbourhood-like region (see Fig.~\ref{fig:splash_feh_vphi}).

In a first step, we replicate the density distribution plots in $\mathrm{[Fe/H]}$ vs. $V_\varphi$ space that \citet[][their Fig.~1]{Belokurov2020} used to identify Splash stars. We show both the row-normalised density distribution in Figs.~\ref{fig:splash_feh_vphi}a and \ref{fig:splash_feh_vphi}b as well as the logarithmic density distributions in Figs.~\ref{fig:splash_feh_vphi}c and \ref{fig:splash_feh_vphi}d. Here we indeed find a strikingly similar distribution in $\mathrm{[Fe/H]}$ vs. $V_\varphi$ space to that observed by \citet{Belokurov2020} for the Solar neighbourhood. In particular, we recover the sickle-like distribution of highest densities (black) around $\mathrm{[Fe/H]} \sim -0.25$ in the row-normalised distribution of Fig.~\ref{fig:splash_feh_vphi}b, which \citet{Belokurov2020} attributed to the three zones of thin disc (top), thick disc (middle), and Splash (bottom).

In the other panels of Fig.~\ref{fig:splash_feh_vphi} with logarithmic density distributions, metal-poor in-situ stars exhibit a broad spread in azimuthal velocity extending from stars with $V_\varphi = -300\,\mathrm{km\,s^{-1}}$ to stars with $+400\,\mathrm{km\,s^{-1}}$, but with a modest prograde bias rather than being centred exactly at $V_\varphi = 0\,\mathrm{km\,s^{-1}}$. For stars with $[\mathrm{Fe/H}] < -0.4$, the median azimuthal velocity is $V_\varphi = 30_{-100}^{+110}\,\mathrm{km\,s^{-1}}$ for the whole galaxy and $V_\varphi = 70_{-70}^{+80}\,\mathrm{km\,s^{-1}}$ within the Solar neighbourhood. Both \citet{Yu2023b} and \citet{Chandra2024} described this as the first of three phases of disc evolution, the chaotic bursty phase. Such prograde behaviour may arise from several processes, including the angular momentum imparted by mergers and gas accretion or the torquing of pre-existing stellar populations during merger events \citep{Santistevan2021}.

As metallicity increases, the stellar density rises and peaks around $-0.4 < \mathrm{[Fe/H]} < 0.1$ for stars with $V_\varphi < 0\,\mathrm{km\,s^{-1}}$. In this metallicity range, we observe a steep increase in the average $V_\varphi$. This second phase of disc evolution \citep{Chandra2024} is also described as spin-up/bursty-disc phase  \citep{Yu2023b, Viswanathan2025b}. Finally, the most metal-rich stars with $\mathrm{[Fe/H]} > 0.1$ are moving on rotationally supported disc orbits with a narrower distribution of $150_{-50}^{+40}\,\mathrm{km\,s^{-1}}$, corresponding to the third, thin-disc phase of disc evolution \citep{Yu2023b, Chandra2024}. Studies of stellar kinematics in cosmological simulations further indicate that much of the velocity dispersion of older stellar populations was already present at formation, with subsequent dynamical heating contributing primarily to the radial component and major mergers influencing the settling of the disc \citep{McCluskey2024}. We note that the hook of stars with $\mathrm{[Fe/H]} > 0.3$ across the galaxy (left panels of Fig.~\ref{fig:splash_feh_vphi}) showing decreasing azimuthal velocities arises because these stars primarily reside in the inner galaxy. In particular, stars with $\mathrm{[Fe/H]} > 0.3$ have a median galactocentric radius of $R = 0.9_{-0.4}^{+0.6}\,\mathrm{kpc}$ in the simulation, indicating that they belong to the central bulge component.

In observations, the interpretation of the [Fe/H]-$V_\varphi$ plane is complicated by uncertainties in stellar ages and the unknown birth locations of stars. In contrast, our simulation allows us to overcome these limitations by providing both precise stellar ages and a clear distinction between in-situ and accreted stars. Taking advantage of this, we define several subsamples of the Solar neighbourhood (Sample~A, Fig.~\ref{fig:splash_feh_vphi}d) to probe the nature of the Splash population. Sample D (Fig.~\ref{fig:splash_feh_vphi}f) selects in-situ Splash-like stars with $-0.4 < \mathrm{[Fe/H]} < 0.1$ and $-100 < V_\varphi < 75\,\mathrm{km\,s^{-1}}$.  More generally, we then define:
\begin{itemize}[leftmargin=2em,labelwidth=2em]
    \item Sample A: all stars in the Solar neighbourhood (Fig.~\ref{fig:splash_feh_vphi}d) \\
    and from within this region we sub-select (Figs.~\ref{fig:splash_feh_vphi}f and \ref{fig:splash_feh_vphi}h):
    \item Sample B: more metal-rich in-situ stars ($\mathrm{[Fe/H]} > 0.1$),
    \item Sample C: in-situ stars with Splash-like [Fe/H] but higher $V_\varphi$,
    \item Sample D: the Splash stars themselves,
    \item Sample E: more metal-poor in-situ stars ($\mathrm{[Fe/H]} < -0.4$), and
    \item Sample F: accreted stars in the Solar neighbourhood.
\end{itemize}

While we identify the Splash-like sample D in Fig.~\ref{fig:splash_feh_vphi} at $-0.4 < \mathrm{[Fe/H]} < 0.1$ in the simulation, we note that \citet{Belokurov2020} found it as slightly lower metallicities of $-0.7 < \mathrm{[Fe/H]} < -0.25$ in the Milky Way's Solar neighbourhood. Because of the findings by \citet{Buck2021}, who studied the influence of chemical evolution model choices, such as nucleosynthetic yields, on chemical evolution predictions on NIHAO-UHD simulations, we attribute this absolute offset in [Fe/H] and other abundances mainly to the choices of chemical evolution parameters in the simulation. Because of these abundance offsets, our selection is consistent with the \textit{relative} position of Splash stars in the [Fe/H] vs. $V_\varphi$ diagrams used by \citet{Belokurov2020}, but not identical.

We also note that \citet{Belokurov2020} did not specifically select in-situ stars for their Splash sample. In the Milky Way analogue, however, we only find stars fitting the Splash selection criteria to be born in-situ (compare Figs.~\ref{fig:splash_feh_vphi}f and \ref{fig:splash_feh_vphi}h in the Splash region). We find the accreted stars of the Milky Way analogue moving with higher azimuthal velocities of $V_\varphi = 100_{-70}^{+80}\,\mathrm{km\,s^{-1}}$%
 within the Solar neighbourhood (Fig.~\ref{fig:splash_feh_vphi}h), as compared to values of $V_\varphi = 20_{-70}^{+100}\,\mathrm{km\,s^{-1}}$ for \textit{GSE} stars within the Milky Way's Solar neighbourhood \citep{Buder2022}. This difference likely reflects the more circular inflow of the simulated merger progenitor, but does not affect the identification of the Splash population in the [Fe/H]-$V_\varphi$ plane.

\begin{figure}
    \centering
    \includegraphics[width=\columnwidth]{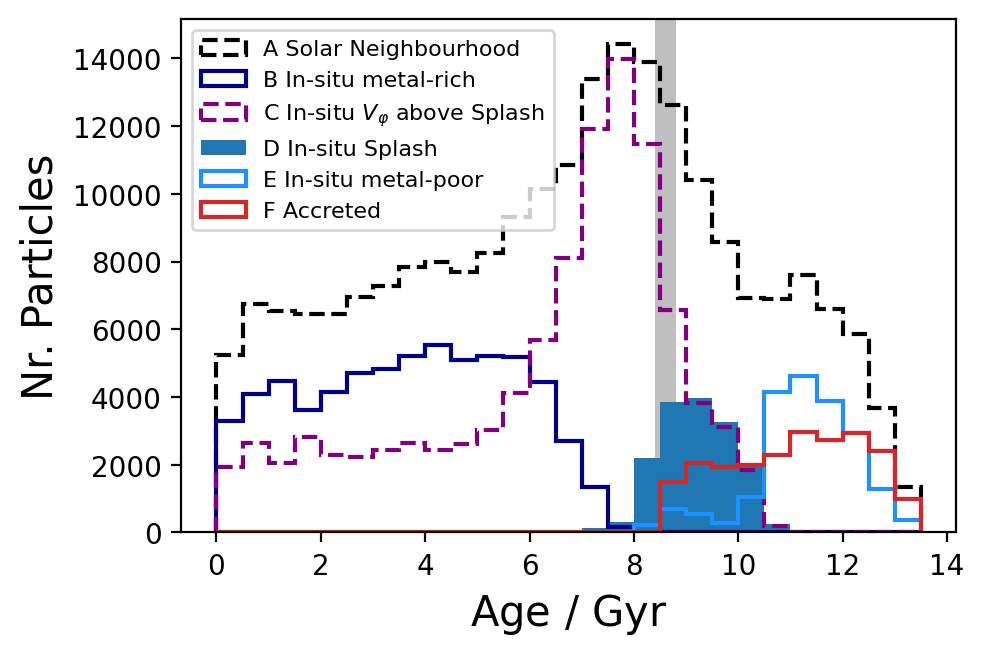}
    \caption{Age distribution of different samples of stars in the Solar neighbourhood as selected in the $\mathrm{[Fe/H]}$ vs. $V_\varphi$ plane of Fig.~\ref{fig:splash_feh_vphi}. A grey bar indicates the time of the major merger around $8.6\,\mathrm{Gyr}$ \href{https://github.com/svenbuder/golden_thread_II/tree/main/figures/splash_age.png}{\faGithub}.}
    \label{fig:splash_age}
\end{figure}

\begin{figure}
    \centering
    \includegraphics[width=\columnwidth]{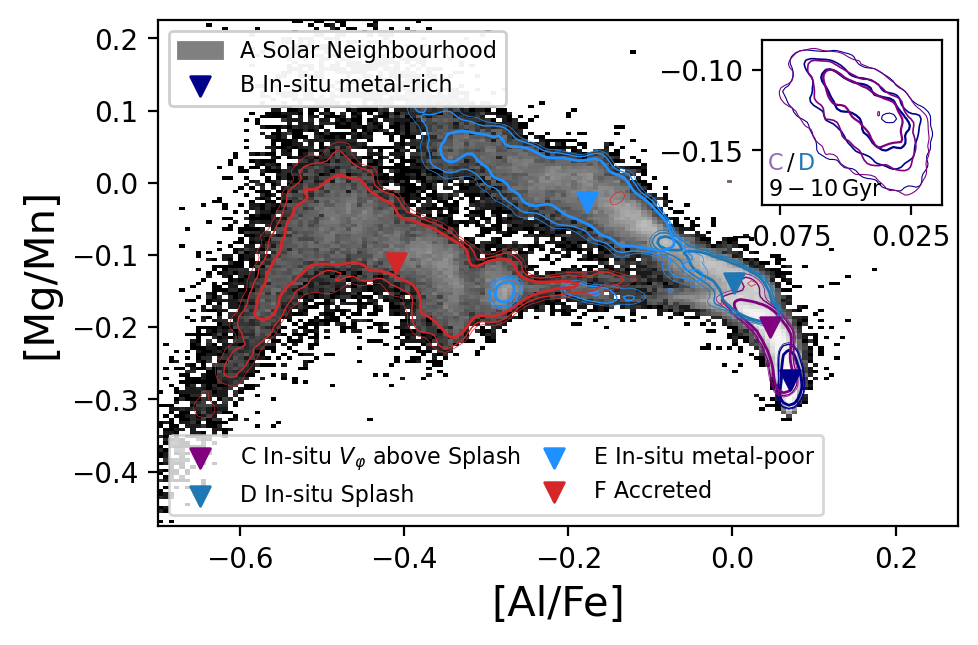}
    \caption{Abundance distribution in [Al/Fe] vs. [Mg/Mn] in different samples of stars in the Solar neighbourhood as selected in the $\mathrm{[Fe/H]}$ vs. $V_\varphi$ plane of Fig.~\ref{fig:splash_feh_vphi}. Contours correspond to the 68\,\% highest-density interval. An inset figure is showing the distribution of samples C and D with ages of $9-10\,\mathrm{Gyr}$, with contours showing the 40, 60, and 80\,\% highest-density intervals \href{https://github.com/svenbuder/golden_thread_II/tree/main/figures/splash_alfe_mgmn_extended.png}{\faGithub}.}
    \label{fig:splash_alfe_mgmn}
\end{figure}

\subsection{Ages and chemistry of Splash-like stars}

Before we take a look at possible changes to the dynamics and spatial distribution of these stars, we get an idea of their typical ages and chemistry in Figs.~\ref{fig:splash_age} and \ref{fig:splash_alfe_mgmn}, respectively. In both spaces, we find the Splash stars as part of the in-situ sequence. This sequence extends from the oldest  ($11.3_{-0.7}^{+0.9}\,\mathrm{Gyr}$%
) and most metal- and [Al/Fe]-poor in-situ stars to younger Splash stars ($9.2_{-0.7}^{+0.7}\,\mathrm{Gyr}$%
) with higher [Al/Fe]. From there, the sequence continues to even younger stars ($7.2_{-3.8}^{+1.3}\,\mathrm{Gyr}$%
), which have similar [Fe/H] but higher $V_\varphi$ and [Al/Fe] and lower [Mg/Mn]. Finally, it reaches the youngest and most metal-rich in-situ stars ($3.8_{-2.5}^{+2.0}\,\mathrm{Gyr}$%
). Furthermore, we note that the accreted stars with ages of $11.2_{-1.7}^{+1.3}\,\mathrm{Gyr}$%
 are for the most part older or at least as old as the Splash stars. This finding (of F older than D older than C) is consistent with the observations by \citet{Belokurov2020}. In particular, we also find that the Splash-like stars are exclusively old, corroborating that they indeed could come from a Splash-like event. We also note that while there are stars with chemical compositions literally bridging the accreted (blue) and in-situ (red) populations, this bridge\footnote{The region of $-0.2 < \mathrm{[Al/Fe]} < -0.05$ and $\mathrm{[Mg/Mn]} < -0.14$.} connects at the low [Mg/Mn]-end of the Splash star region. As already noted by \citet{Buder2024}, these stars  (less than 1\,\% within the Solar neighbourhood) are likely formed from mixed gas during the merger, but do not contribute significantly (less than 9\,\%) to any of the sub-samples of the simulation's Solar neighbourhood.

More strikingly, we note only a little overlap in the distribution of ages and chemistry of the most metal-poor in-situ stars with the Splash stars within the Solar neighbourhood. The age distribution of the Splash stars (Sample D) is consistent with the oldest stars of the in-situ population with similar [Fe/H] but higher $V_\varphi$ than the Splash stars (Sample C) and shows a similar profile especially for ages of $9-10\,\mathrm{Gyr}$.

This result opens up a number of questions that evolve around how special the Splash stars are -- both in terms of their chemistry as well as their orbits. While \citet{Belokurov2020} favoured a scenario in which the stars have been splashed by a major merger, they also discussed a scenario in which the stars might have been born dynamically hotter with more eccentric orbits. One could for example imagine these stars being the low azimuthal velocity tail of a dynamically hotter early disc. We thus take a closer look at the sample of stars with similar iron abundance ($-0.4 < \mathrm{[Fe/H]} < 0.1$) and the aforementioned ages of $9-10\,\mathrm{Gyr}$.

\begin{figure}
    \centering
    \includegraphics[width=\columnwidth]{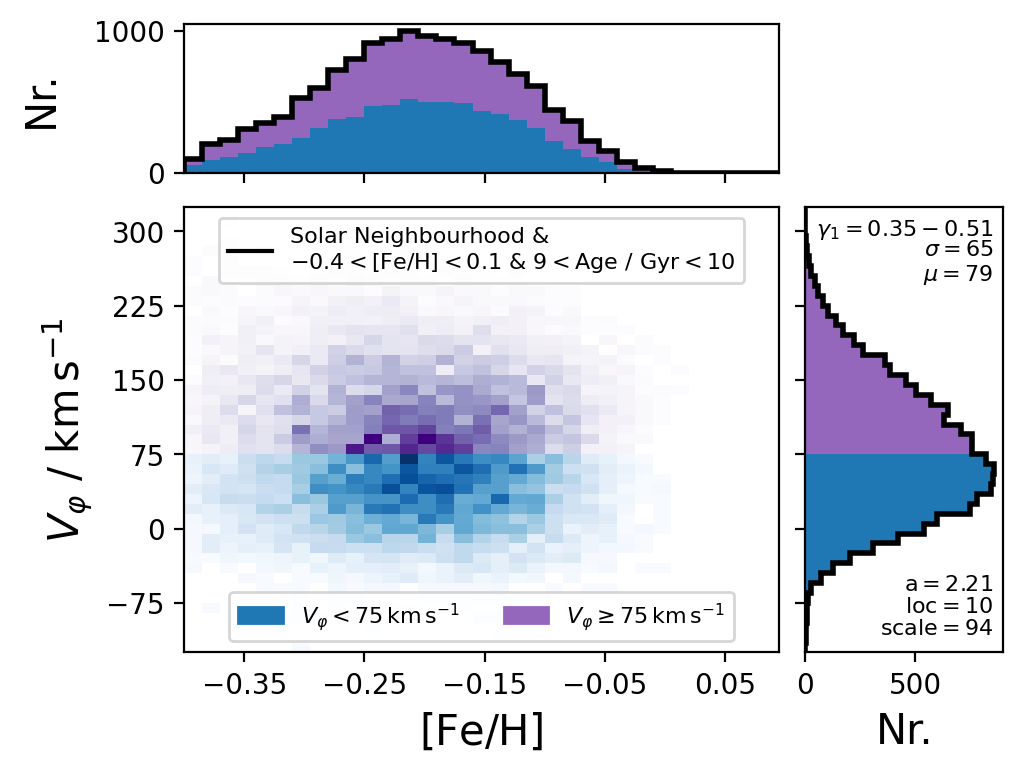}
    \caption{Distributions of [Fe/H] and $V_\varphi$ for stars in the Solar neighbourhood at $-0.4 < \mathrm{[Fe/H]} < 0.1$. We show the stacked distribution (black lines) as well as 2-dimensional and 1-dimensional histograms of each distribution.
    \href{https://github.com/svenbuder/golden_thread_II/tree/main/figures/splash_vphi_distribution.png}{\faGithub}.}
    \label{fig:splash_vphi_distribution}
\end{figure}

\begin{figure*}
    \centering
    \includegraphics[width=0.995\textwidth]{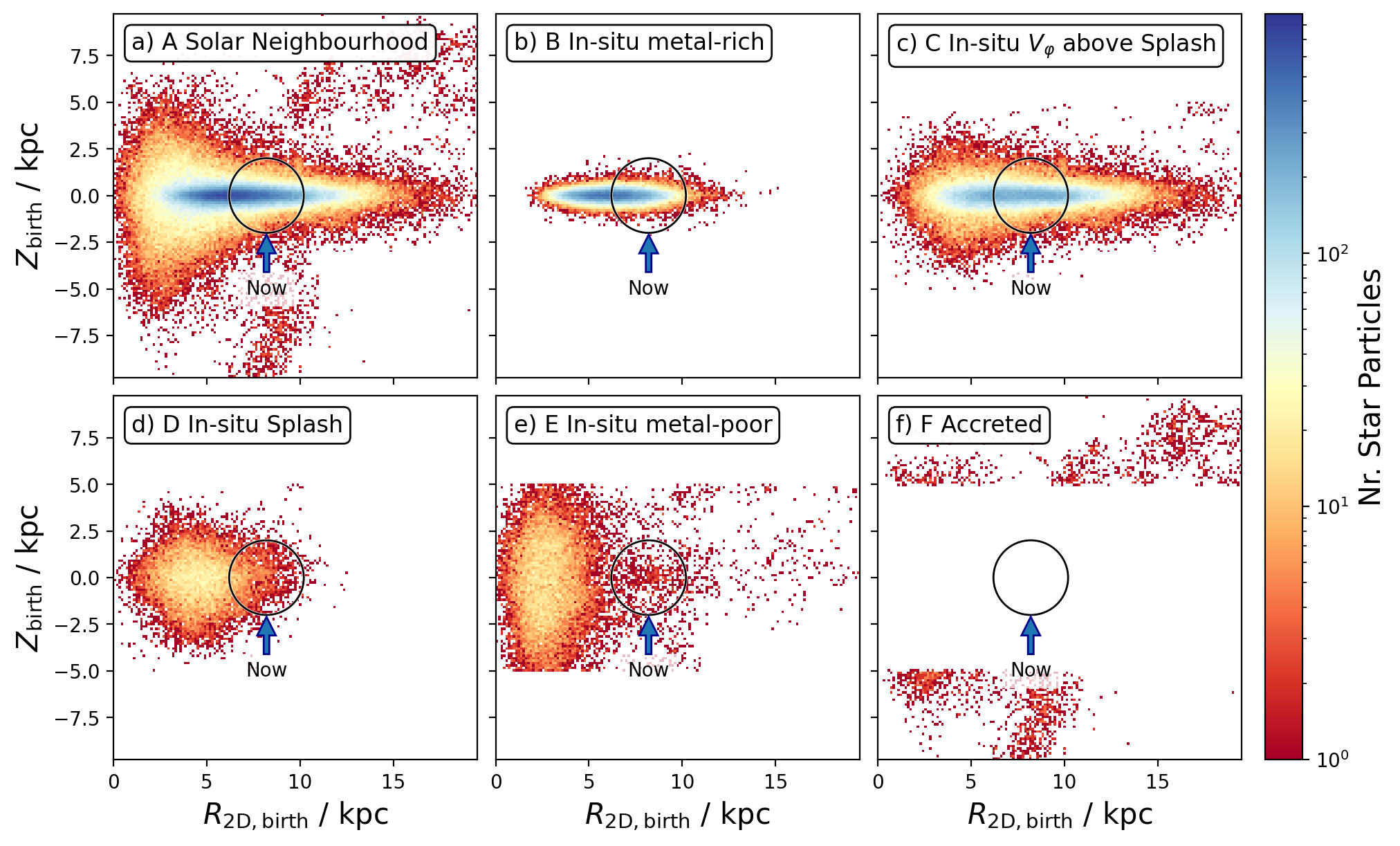}
    \caption{Density distribution of birth positions in galactocentric cylindrical coordinates $R_\mathrm{birth, 2D}$ and $Z_\mathrm{birth}$ for star particles of samples A-F (corresponding to panels a-f) that are currently in the Solar neighbourhood (black circles) of $2\,\mathrm{kpc}$ around $R_\mathrm{2D} = 8.2\,\mathrm{kpc}$. In panels e) and f) we note the imprint of our selection of in-situ vs. accreted stars via $\vert Z_\mathrm{birth} \vert > 5\,\mathrm{kpc}$ \citepalias[Eqs.~4 and 5 of][]{Buder2025c} \href{https://github.com/svenbuder/golden_thread_II/tree/main/figures/splash_rbirth_zbirth.png}{\faGithub}.}
    \label{fig:splash_rbirth_zbirth}
\end{figure*}

\begin{figure}
    \centering
    \includegraphics[width=\columnwidth]{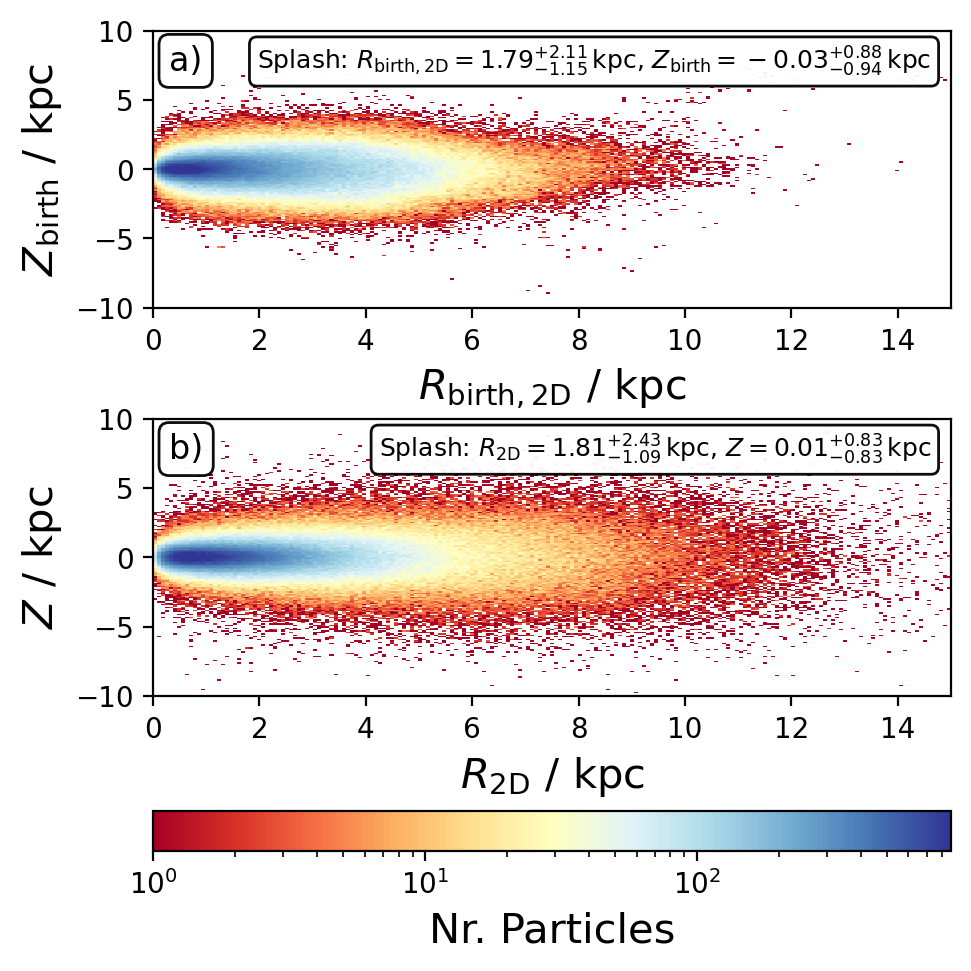}
    \caption{Radial and vertical distribution of Splash stars. Panel a) shows birth radii $R_\mathrm{birth, 2D}$ and birth heights $Z_\mathrm{birth}$, whereas panel b) shows current radii $R_\mathrm{2D}$ and heights $Z$. Text insets show the median and 16th to 84th percentiles of each distribution
    \href{https://github.com/svenbuder/golden_thread_II/tree/main/figures/splash_rz_birth_now.png}{\faGithub}. We note that only 3.1\% and 7.7\% of Splash stars reside beyond $R_\mathrm{2D} > 6\,\mathrm{kpc}$ at birth and at present, respectively.}
    \label{fig:splash_rz_birth_now}
\end{figure}

\subsection{Coeval counterparts of Splash-like stars} \label{sec:vphi_skewness}

In particular, we are interested in a comparison of properties of the low azimuthal velocity (Splash) stars with $V_\varphi < 75\,\mathrm{km\,s^{-1}}$ and their higher azimuthal velocity counterpart. We show their distribution in a [Fe/H] vs. $V_\varphi$ diagram in Fig.~\ref{fig:splash_vphi_distribution} and notice that their distribution both in 2-dimensions as well as each individual dimension appear rather smooth. In particular the distribution of azimuthal velocities, $V_\varphi = 73_{-59}^{+74}\,\mathrm{km\,s^{-1}}$%
, appears normal-distributed with a slight positive skewness. We have thus used the \textsc{skewnorm} functionality of the \textsc{scipy.stats} module \citep{Scipy} to fit a skewed normal distribution to the data, finding a distribution around $79 \pm 65\,\mathrm{km\,s^{-1}}$ with a positive skewness of $\gamma_1$ between 0.35 and 0.51 when using \textsc{skew} or \textsc{skewnorm}, respectively. In addition, we have also looked into the chemical enrichment of the low- and high-$V_\varphi$ samples, in particular [Al/Fe] and [Mg/Mn]. The samples are visually congruent with each other (see inset of Fig.~\ref{fig:splash_alfe_mgmn}) and distributed in the region of Fig.~\ref{fig:splash_alfe_mgmn} of sample D. To statistically assess whether the two 2-dimensional distributions are drawn from the same underlying population, we apply the Maximum Mean Discrepancy test \citep{Gretton2012} using the \textsc{hyppo} package's \textsc{ksample.MMD} \citep{hyppo}. The resulting test statistic of $10^{-6}$ with a $p$-value of 0.26 indicates no significant difference between the two samples. For reference, Kolmogorov–Smirnov tests \citep{Hodges1958} applied independently to the 1D marginals via \textsc{scipy.stats.ks\_2samp} also yield high $p$-values of 1.0 and 0.86. These results indicate that the low-$V_\varphi$ and high-$V_\varphi$ Splash-like stars are consistent with being drawn from the same underlying population with a broad spread in azimuthal velocity. In this work, the characteristic shape of the $\mathrm{[Fe/H]}$-$V_\varphi$ distribution is therefore interpreted as the primary observable signature of the Splash population.

In the original interpretation of the Splash population by \citet{DiMatteo2019} and \citet{Belokurov2020}, stars of the early disc were thought to have been scattered onto halo-like orbits, which would correspond to a skewing toward lower angular momentum (i.e. lower $V_\varphi$). In contrast, the Splash-like population in our simulation shows a skew toward positive $V_\varphi$. This difference highlights a tension between the expected kinematic signature of a splashing event and the distribution produced in this simulation. The origin of this discrepancy may lie either in observational selection effects when identifying Splash stars in the Milky Way, or in differences in the merger geometry and angular momentum of the simulated interaction. We discuss these possibilities further in Section~\ref{sec:discussion}.

\begin{figure*}
    \centering
    \includegraphics[width=\textwidth]{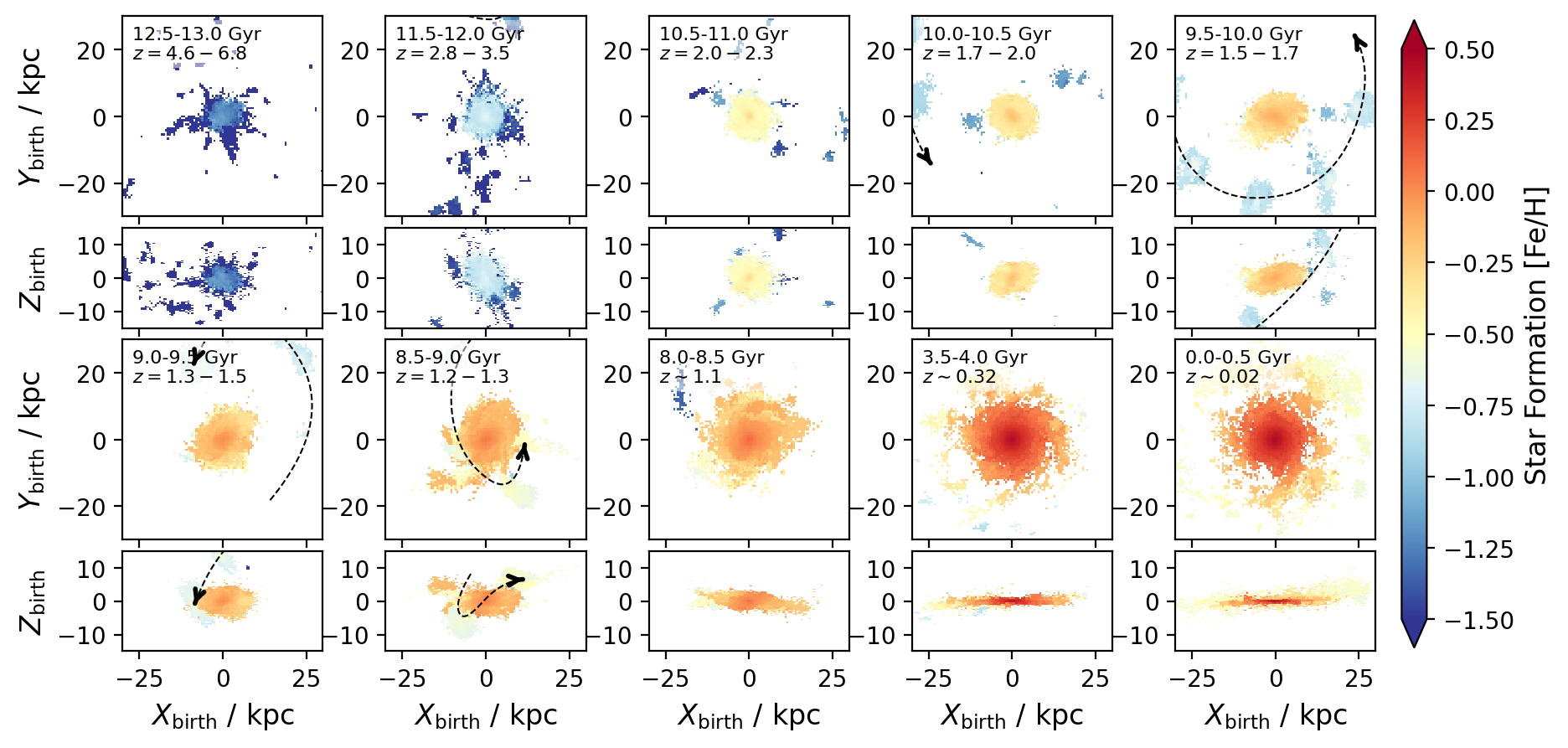}
    \caption{Distributions of iron abundances {[Fe/H]} for the face-on ($X_\mathrm{birth}$ vs. $Y_\mathrm{birth}$, odd rows) and edge-on ($X_\mathrm{birth}$ vs. $Z_\mathrm{birth}$, even rows) views in ten selected birth epochs within $0.5\,\mathrm{Gyr}$, starting from $12.5-13.0\,\mathrm{Gyr}$ in the top-left panel to $0.0-0.5\,\mathrm{Gyr}$ in the bottom-right panel. Dashed black arrows indicate the path of the last major merger. See \href{https://github.com/svenbuder/golden_thread_II/tree/main/figures}{\faGithub} for figures with all epochs as well as colouring by density \href{https://github.com/svenbuder/golden_thread_II/tree/main/figures/trace_star_formation_xy_xz_feh_selection.png}{\faGithub}.}
    \label{fig:trace_star_formation_xy_xz_feh_selection}
\end{figure*}

\subsection{Birth and present-day orbits of Splash-like stars}

To further establish, where -- and by proxy on which orbits -- the Splash stars were born, we analyse their and the other samples' birth positions in Fig.~\ref{fig:splash_rbirth_zbirth}. The simulation makes remarkable predictions about the origin of stars that are currently residing in a small Solar neighbourhood-like region, with stars visiting from the centre of the galaxy or its furthest outskirts at $20\,\mathrm{kpc}$ (see Fig.~\ref{fig:splash_rbirth_zbirth}a) -- not to mention the accreted stars born even further away (Fig.~\ref{fig:splash_rbirth_zbirth}f). In agreement with radial migration models \citep[for example][]{Frankel2018, Frankel2020}, a significant amount of metal-rich stars (Fig.~\ref{fig:splash_rbirth_zbirth}b) is visiting from the inner galaxy (Sample B with 5-95th percentiles of $R_\mathrm{birth, 2D} \sim 3.9 - 8.9\,\mathrm{kpc}$ centred around a median of $6.3\,\mathrm{kpc}$), while high-$V_\varphi$ stars with Solar-like iron abundances (Sample C) show a larger distribution of birth origins along the disc ($R_\mathrm{birth, 2D} \sim 3.5-12.5\,\mathrm{kpc}$, centred around $7.4\,\mathrm{kpc}$). However, both the in-situ Splash ($R_\mathrm{birth, 2D} \sim 2.2-7.8\,\mathrm{kpc}$, centred around $4.6\,\mathrm{kpc}$) and the in-situ metal-poor stars ($R_\mathrm{birth, 2D} \sim 1.0-9.7\,\mathrm{kpc}$, centred around $2.9\,\mathrm{kpc}$) are basically born exclusively in the inner galaxy ($R_\mathrm{birth, 2D} \ll 8.2\,\mathrm{kpc}$) and with a larger spread of $Z_\mathrm{birth}$, that is, not on a thinner disc.

Since the Splash stars were born further inwards, the question arises, whether they were born on more radially eccentric orbits which reach into our Solar neighbourhood, or if their orbits have significantly changed through the radial splashing that shifts their initially inner orbits to overlap with our Solar neighbourhood. Because our snapshot at redshift $z = 0$ only includes the present-day velocities, we turn to a comparison of birth and present-day spatial distribution in Fig.~\ref{fig:splash_rz_birth_now} in an effort to quantify the change of spatial distributions which would imply a change of orbits. When comparing the birth (upper panel) with present-day (lower panel) density distributions in linear density, we note a more extended low-density envelope to higher radii. Similarly, we note an extend of larger $Z$-coverage for $R_\mathrm{2D} > 6\,\mathrm{kpc}$ -- again with very low densities. The overall distribution of stars has, however, changed insignificantly. This is evident in the only minor change of the average $R_\mathrm{birth,2D} = 1.79_{-1.15}^{+2.11}\,\mathrm{kpc}$%
 to $R_\mathrm{2D} = 1.81_{-1.09}^{+2.43}\,\mathrm{kpc}$%
 and $Z_\mathrm{birth} = -0.03_{-0.94}^{+0.88}\,\mathrm{kpc}$%
 to $Z = 0.01_{-0.83}^{+0.83}\,\mathrm{kpc}$%
. When calculating the average of the change for the individual stars, we find $\overline{\Delta R_\mathrm{2D}} = \overline{R_\mathrm{2D} - R_\mathrm{2D,birth}} = 0.1_{-1.1}^{+1.2}\,\mathrm{kpc}$%
 and $\overline{\Delta |Z|} = \overline{|Z| - |Z_\mathrm{birth}|} = -0.03_{-0.70}^{+0.68}\,\mathrm{kpc}$%
. In our particular Milky Way analogue simulation, this thus implies little orbit change of these stars, or no significant splashing, with a possible exception of stars beyond $R_\mathrm{2D} > 3.9\,\mathrm{kpc}$, the outermost $16\,\mathrm{\%}$ of Splash stars (the 84th percentile of $R_\mathrm{birth,2D}$). We discuss the implications of this finding in Section~\ref{sec:discussion}.

\subsection{Effect on newly forming stars}

Although the main focus of our study is to understand the properties and origin of the Splash, we want to briefly investigate the potential of the available birth positions to study the evolution of the main galaxy's density and chemical enrichment during and after the merging. For this purpose, we investigate the change of iron abundance [Fe/H] in birth positions in both the $X_\mathrm{birth}$ vs. $Y_\mathrm{birth}$ and $X_\mathrm{birth}$ vs. $Z_\mathrm{birth}$ dimensions in Fig.~\ref{fig:trace_star_formation_xy_xz_feh_selection}, with each panel showing the median iron abundance in each bin for stars formed within a selection of 10 star formation epochs that each span $0.5\,\mathrm{Gyr}$. We remind ourselves that because birth positions are estimated in $0.1\,\mathrm{Gyr}$ steps, we expect the footprint of the major accretion to mimic up to five smaller galaxies, which in fact are showing the position of the same smaller galaxy in $0.1\,\mathrm{Gyr}$ steps (see especially the panels around $9.5-10.0\,\mathrm{Gyr}$ in these figures).

From these figures we can assess that the stars of the early main galaxy were born on typically random orbits leading to a spheroidal galaxy shape and new stars following similar orbits until a lookback time of $\sim 10\,\mathrm{Gyr}$. Between $\sim10$ and $\sim 8.5\,\mathrm{Gyr}$ ago, we then see the birth positions follow a more and more disc-like structure with increasing radii and decreasing heights. For ages below $8\,\mathrm{Gyr}$, we then see star formation occurring within a rotationally supported disc. These observations are in agreement with the modelling by \citet{MCM2013}. 
Albeit qualitatively in agreement with the description of the \textit{spin-up} by \citet{Belokurov2022}, we would place the tracers of the spin-up phase exactly at the position of the Splash in the $\mathrm{[Fe/H]}$ vs. $V_\varphi$ \citep[see also][]{Viswanathan2025b}. In contrast to \citet[][their Fig.~5b]{Belokurov2022}, who associate the culmination of the spin-up phase with stars at $\mathrm{[Fe/H]} \sim -0.9$, our simulation indicates that spin-up continues to higher metallicities, up to $\mathrm{[Fe/H]} \lesssim -0.1$, although these bins are increasingly dominated by stars already born on rotating disc orbits. A detailed assessment of how the smaller galaxy’s gravitational influence and angular momentum transfer between $\sim10$ and $\sim8.6\,\mathrm{Gyr}$ ago contributed to the spin-up of the Milky Way analogue’s disc requires tracing gas and dark matter in addition to stars, which we leave to future work.

What we can investigate, however, is the change of star formation [Fe/H] during this time. This property is used as colour-coding for the different star formation epochs in Fig.~\ref{fig:trace_star_formation_xy_xz_feh_selection}. Here we see that after the initial main galaxy started forming stars with $\mathrm{[Fe/H]} \leq -1.5$ (dark blue), its core quickly started forming more metal-rich stars with the star formation $\mathrm{[Fe/H]}$ in the core rising from $-1.0$ to $-0.75$ between $\sim 12.5$ and $\sim11.5\,\mathrm{Gyr}$ ago, while several smaller amounts of metal-poor stars ($\mathrm{[Fe/H]} \leq -1.5$) where born outside of the core galaxy and quickly accreted. This process continues with star formation in the core reaching $\mathrm{[Fe/H]} \sim -0.2$ and the outskirts reaching $-0.35\,\mathrm{dex}$ at a lookback time of $\sim 10\,\mathrm{Gyr}$. At this time, we notice the arrival of the major merger galaxy to within $30\,\mathrm{kpc}$ distance. Its path is indicated with dashed arrows as it continues to form stars in each $0.1\,\mathrm{Gyr}$ snapshot and finally merge with the main galaxy in a spiral pattern around $8.6\,\mathrm{Gyr}$ ago. For this galaxy, we also note a negative metallicity gradient from core to outskirts, but at lower metallicities than the main galaxy, in particular in the top-right panel of Fig.~\ref{fig:trace_star_formation_xy_xz_feh_selection}. Such gradients were also found by other simulation studies \citep{Amarante2022, Khoperskov2023d, Mori2024, Carrillo2026}. Most importantly, we notice that while the metallicity of the major merger galaxy tends to still increase and reach $\mathrm{[Fe/H]} \sim -0.5$ until $\sim 8.5\,\mathrm{Gyr}$ ago, the main galaxy's star formation [Fe/H] seems to not decrease within the inner $R_\mathrm{birth,2D} < 10\,\mathrm{kpc}$ (compare the snapshots of $8.5-9.0\,\mathrm{Gyr}$ with $8.0-8.5\,\mathrm{Gyr}$ for that region). Only in the outskirts of the galaxy at $10 \lesssim R_\mathrm{birth, 2D} \lesssim 20\,\mathrm{kpc}$ do we note some patches of lower star formation [Fe/H]. We note that \citet{Buck2023} found considerable dips in the gas metallicity [Fe/H] of this simulation during the merger. By comparison with the positions of the major merger galaxy around this time (see Fig.~9b from \citetalias{Buder2025c}), we can confirm those as accreted structures as well. This again underlines our impression of no significant lowering of star formation [Fe/H] during the merger. This behaviour is consistent with recent results from cosmological simulations, which show that mergers alone do not necessarily induce strong or sustained metal dilution of the star-forming gas, and that the evolution of chemical sequences is instead governed by a combination of gas accretion and star-formation history \citep{Parul2025, Orkney2026}. For all remaining epochs, we only notice a rather uneventful inside-out star formation with a radial metallicity gradient \citep[see also][]{Buder2025}.

\section{Discussion}
\label{sec:discussion}

In \citetalias{Buder2025c}, we examined the efficiency of recovering accreted stars in integral-of-motion space and assessed how much information about their birth positions and chemistry is preserved. Here, we extend this framework to focus on what can be inferred from birth positions, ages, and chemistry about the dynamical state of the early disc, and in particular on the proposed splashing of the Milky Way's protodisc \citep{Belokurov2020}.

\subsection{The nature of Splash-like stars}

Our analysis of Splash-like stars in a Solar-like neighbourhood with low azimuthal velocities and subsolar metallicities (Figs.~\ref{fig:splash_feh_vphi}--\ref{fig:splash_rz_birth_now}) suggests that these stars represent the ancient portion of the protodisc, in line with earlier observational studies \citep{Bonaca2017, Haywood2018b, DiMatteo2019, Gallart2019, Belokurov2020, Viswanathan2025b}. Similar to \citet{Belokurov2020}, we find that the properties of these stars bridge smoothly to those of the thick disc. However, when comparing stars with similar chemistry and ages, we also identify prograde counterparts with high $V_\varphi$, and that the $V_\varphi$ distribution of the sample shows a positive skewness (Fig.~\ref{fig:splash_vphi_distribution}). This indicates that the majority of stars in this population already occupy lower $V_\varphi$ orbits, rather than being drawn from a dynamically cold distribution that was later heated. The latter would lead to a negative skewness with a higher mean $V_\varphi$, which is inconsistent with our findings.

As discussed in Section~\ref{sec:vphi_skewness}, the low- and high-$V_\varphi$ Splash-like stars are consistent with being drawn from the same underlying population, suggesting that the present-day spread in azimuthal velocity reflects a broad kinematic distribution rather than two distinct stellar components.
This behaviour differs from the original interpretation of the Splash population by \citet{Belokurov2020}, where stars were thought to have been scattered from a dynamically colder proto-disc onto halo-like orbits. Such a process would naturally produce a distribution skewed toward lower angular momentum, that is, lower $V_\varphi$. In contrast, the simulation analysed here exhibits a positive skewness in $V_\varphi$, indicating a different kinematic imprint.

As noted in Section~\ref{sec:analysis_infall}, the last major merger in this simulation follows a relatively circular infall trajectory, in contrast to the predominantly radial orbit inferred for the Milky Way's \textit{GSE} event. Previous simulation studies have shown that the efficiency of disc heating and the resulting fraction of Splash-like stars depend sensitively on the merger properties, including mass ratio, orbital eccentricity, and the dynamical state of the proto-disc \citep[e.g.][]{Grand2020}. Radial mergers tend to produce stronger impulsive heating and larger Splash fractions, whereas more circular mergers redistribute angular momentum over longer timescales through gravitational torques. In such cases, the heated stellar population may retain a net prograde motion, naturally producing the positive skewness in $V_\varphi$ seen in our simulation. We have therefore traced the time evolution of the $V_\varphi$ as well as the $\mathrm{[Fe/H]}$-$V_\varphi$ distribution before, during, and after the merger interaction, which is shown in Appendix~\ref{sec:appendix_A}. We find that while a subset of stars, particularly those that end up in the Solar neighbourhood, exhibits a measurable evolution toward lower angular momentum during the merger, the bulk of the Splash-like population retains a similar kinematic distribution, indicating that the merger does not uniformly reshape the full population.

These results highlight a tension in the interpretation of Splash-like populations: either observational selections in the Milky Way preferentially trace the low-$V_\varphi$ tail of a broader in-situ population, or the kinematic imprint of the Splash depends sensitively on the merger geometry. We therefore caution that the characteristic $\mathrm{[Fe/H]}$-$V_\varphi$ structure alone may not uniquely diagnose the underlying formation pathway.

Our findings do not exclude merger-driven heating as an origin for Splash-like stars. However, they indicate that such a pattern may not only arise from the heating of a dynamically cold proto-disc alone. Instead, the results are consistent with a scenario in which at least part of the population was already dynamically hot prior to the merger, with the interaction further redistributing angular momentum.

This challenges the interpretation that the Splash is simply the tail of a cold protodisc violently heated by \textit{GSE} \citep{DiMatteo2019, Belokurov2020}. Instead, our findings align more closely with the results of \citet{Amarante2020}, who showed in an isolated clumpy disc simulation that stars can be \emph{born hot} on halo-like orbits while retaining thick-disc chemistry. Our analysis of birth positions strengthens this view: we find no significant radial or vertical displacement between average present-day and birth positions ($\overline{\Delta R_\mathrm{2D}} = \overline{R_\mathrm{2D} - R_\mathrm{2D,birth}} = 0.1_{-1.1}^{+1.2}\,\mathrm{kpc}$%
, $\overline{\Delta |Z|} = \overline{|Z| - |Z_\mathrm{birth}|} = -0.03_{-0.70}^{+0.68}\,\mathrm{kpc}$%
), suggesting that a large-scale dynamical splashing event may not be required to explain the full population of Splash-like stars, and that at least part of this population may already have been dynamically hot prior to the merger.

\subsection{Implications for disc heating and angular momentum build-up}

The galaxy in which Splash stars formed was not yet rotation-dominated. In the simulation, a rotation-supported disc emerges only $\sim 9.5\,\mathrm{Gyr}$ ago, broadly consistent with observational inferences of a gradual increase in disc rotation from 11 to 8 Gyr ago \citep[e.g.][]{MCM2013}. We note, however, that the timing of disc spin-up inferred from present-day kinematics may not directly reflect the true evolutionary history, as merger-driven perturbations can significantly reshape stellar orbits over time \citep[e.g.][]{Orkney2026b}. Whether this angular momentum build-up was causally linked to the last major merger remains uncertain. While some studies argue that \textit{GSE} heating produced the Splash \citep{Bonaca2020, DiMatteo2019}, our results suggest that at least part of the population was already dynamically hot. The reality may well involve a combination of both effects, as also hinted by cosmological simulations that show Splash-like stars arising in multiple pathways, including major mergers, retrograde minor mergers, and secular processes \citep{Quinn1993, Purcell2010, Villalobos2008, Villalobos2009, Dillamore2023, Dillamore2025, Kisku2025}. While secular processes such as bar resonances have also been shown as a possible pathway to generate Splash-like signatures \citep{Dillamore2022,Dillamore2023}, we note that the NIHAO-UHD simulation analysed here does not form a significant present-day bar \citep{Buder2025}. This makes it unlikely that bar-driven resonances are responsible for the Splash-like stars we identify at $z=0$, at least in this particular simulation.

Quantifying the merger contribution to angular momentum evolution is a broader challenge. The outcome depends on mass ratios, orbital configurations, and the timing of gas inflows \citep[e.g.][]{Lagos2017, Lagos2018}. A single simulation cannot establish causality, but suites of cosmological simulations offer promising routes forward \citep{Fattahi2019, Pillepich2019, SotilloRamos2022}. Our findings underscore the need for such systematic studies to disentangle the relative importance of heating versus hot-born origins.

\subsection{Limitations and outlook}

In this work our analysis focused solely on the stellar component, in line with current observations. We do not explicitly follow gas inflows onto the Milky Way analogue or the gas content of the Milky Way analogue and its major merger, which are known to influence the degree of mixing and subsequent star formation \citep[e.g.][]{Cooper2015, Agertz2021, Renaud2021b, Buck2023, TepperGarcia2024}. Future work combining stellar and gaseous tracers may better constrain the role of gas-rich accretion in producing stars with Splash-like properties.

As more precise observational constraints become available, particularly in ages and multi-element abundance patterns, it should become possible to further test whether Splash-like stars were primarily heated, born hot, or both. The emerging diversity of predictions across simulation suites \citep{Brooks2026} highlights the value of confronting these with observationally defined populations. Our results demonstrate that cosmological simulations with traced birth positions provide a crucial tool for this task, moving beyond kinematics alone and towards a more nuanced understanding of the Splash as a window into our Galaxy's early disc and merger history.

\section{Conclusions}
\label{sec:conclusions}

In \citetalias{Buder2025c}, we established the framework for interpreting the last major merger of a Milky Way analogue, focusing on selection efficiency and the retention of chemodynamical memory. Here, we have used the same NIHAO-UHD simulation to examine the proposed Splash population, enabled by tracing the birth positions, ages, and chemistry of stars.

Our analysis finds no evidence for a dominant large-scale dynamical splashing of in-situ stars. Instead, protodisc stars -- including those with present-day low azimuthal velocities -- were already born on dynamically hot orbits in a turbulent early disc. The transition to a rotation-supported disc occurs only during or after the merger. This suggests that much of the observed Splash may not require violent heating by the \textit{GSE} merger, but can arise naturally from stars formed in a thick, kinematically hot disc, later intermixed with accreted stars and stars formed from merger-driven gas inflows. A subset of stars, particularly those now found in the Solar neighbourhood, does exhibit signatures of merger-driven angular-momentum redistribution, indicating that dynamical heating contributes locally but does not dominate the global population.

These results place our simulation in closer agreement with models in which Splash-like populations are \emph{born hot} in a turbulent disc \citep[e.g.][]{Amarante2020}, during the spin-up phase \citep[e.g.][]{Yu2023}, possibly shaped by feedback-driven \textit{baryon sloshing} in gas-rich progenitors \citep{BlandHawthorn2025}, while still acknowledging that merger-induced heating contributes for a subset of stars, particularly in spatially localised regions, as argued by other studies \citep[e.g.][]{DiMatteo2019, Belokurov2020, Bonaca2020}. Together with recent evidence from cosmological simulations and observations \citep{Dillamore2022, Dillamore2023, Khoperskov2023, Dillamore2025, Kisku2025}, this highlights that Splash-like stars likely arise through a combination of pathways rather than a single mechanism.

We conclude that the interpretation of the Splash as a distinct or predominantly merger-heated population is overly restrictive. A more nuanced view is required, in which kinematics, chemistry, and formation history jointly contribute to the observed distribution of stars. Cosmological simulations with birth-position information provide a critical benchmark for disentangling these processes, and future simulation suites will be essential for quantifying the balance between heated and born-hot origins in the context of the Milky Way’s early assembly.

\section*{Acknowledgments}

We acknowledge the traditional owners of the land on which the ANU stands, the Ngunnawal and Ngambri people. We pay our respects to elders past, and present and are proud to continue their tradition of surveying the night sky and its mysteries to better understand our Universe. SB acknowledges support from the Australian Research Council under grant number DE240100150.
TB's contribution to this project was made possible by funding from the Carl Zeiss Stiftung. \'{A}S~acknowledges funding from the European Research Council (ERC) under the European Union’s Horizon 2020 research and innovation programme (grant agreement No. 101117455).

\section*{Data Availability}

All code to reproduce the analysis and figures can be publicly accessed via \url{https://github.com/svenbuder/golden_thread_II}.
The used simulation snapshot can be publicly accessed as FITS file via \url{https://github.com/svenbuder/preparing_NIHAO}. Original data, more snapshots and other galaxies can be found at \url{https://tobias-buck.de/#sim_data}. We encourage interested readers to get in contact with the authors for full data access and advice for use and cite \citet{Buck2020b, Buck2021}.

\section*{Software}

The research for this publication was coded in \textsc{python} (version 3.12.11) and included its packages
\textsc{astropy} \citep[v. 7.1.0;][]{Robitaille2013,PriceWhelan2018},
\textsc{hyppo} \citep[v. 0.5.2;][]{hyppo},
\textsc{IPython} \citep[v. 9.1.0;][]{ipython},
\textsc{matplotlib} \citep[v. 3.10.3;][]{matplotlib},
\textsc{NumPy} \citep[v. 2.2.6;][]{numpy}, as well as
\textsc{scipy} \citep[v. 1.16.0;][]{Scipy}.
We used \textsc{topcat} \citep[v. 4.7;][]{Taylor2005}.

\bibliographystyle{mnras}
\bibliography{bib}

\appendix

\section{Evolution of the $V_\varphi$ and $\mathrm{[Fe/H]}$-$V_\varphi$ distributions} \label{sec:appendix_A}

\begin{figure}
\centering
\includegraphics[width=\columnwidth]{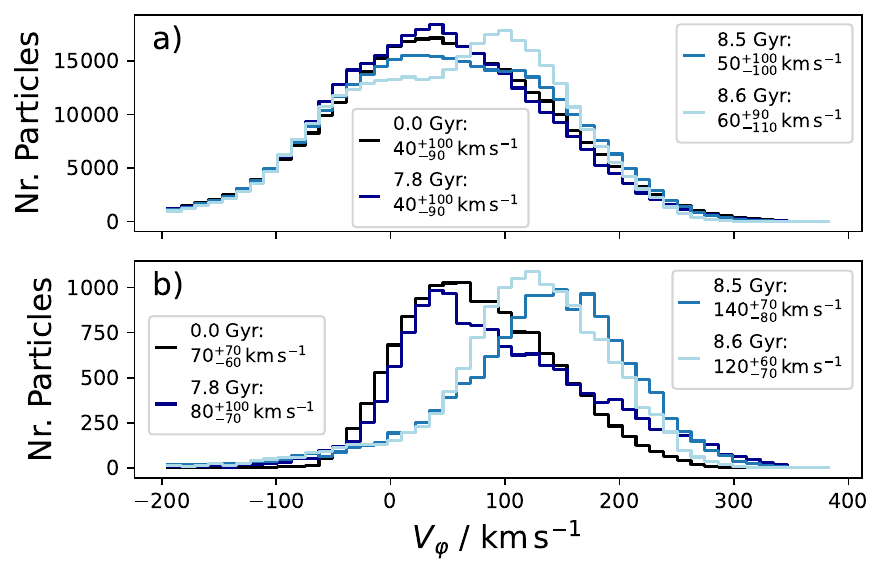}
\caption{
Evolution of the azimuthal velocity distribution of Splash-like stars before, during (10.0-8.6 Gyr ago), and after the merger interaction in a) the full population; and b) stars that end up in the Solar-like neighbourhood at $z=0$.}
\label{fig:splash_vphi_histograms}
\end{figure}

\begin{figure*}
\centering
\includegraphics[width=\columnwidth]{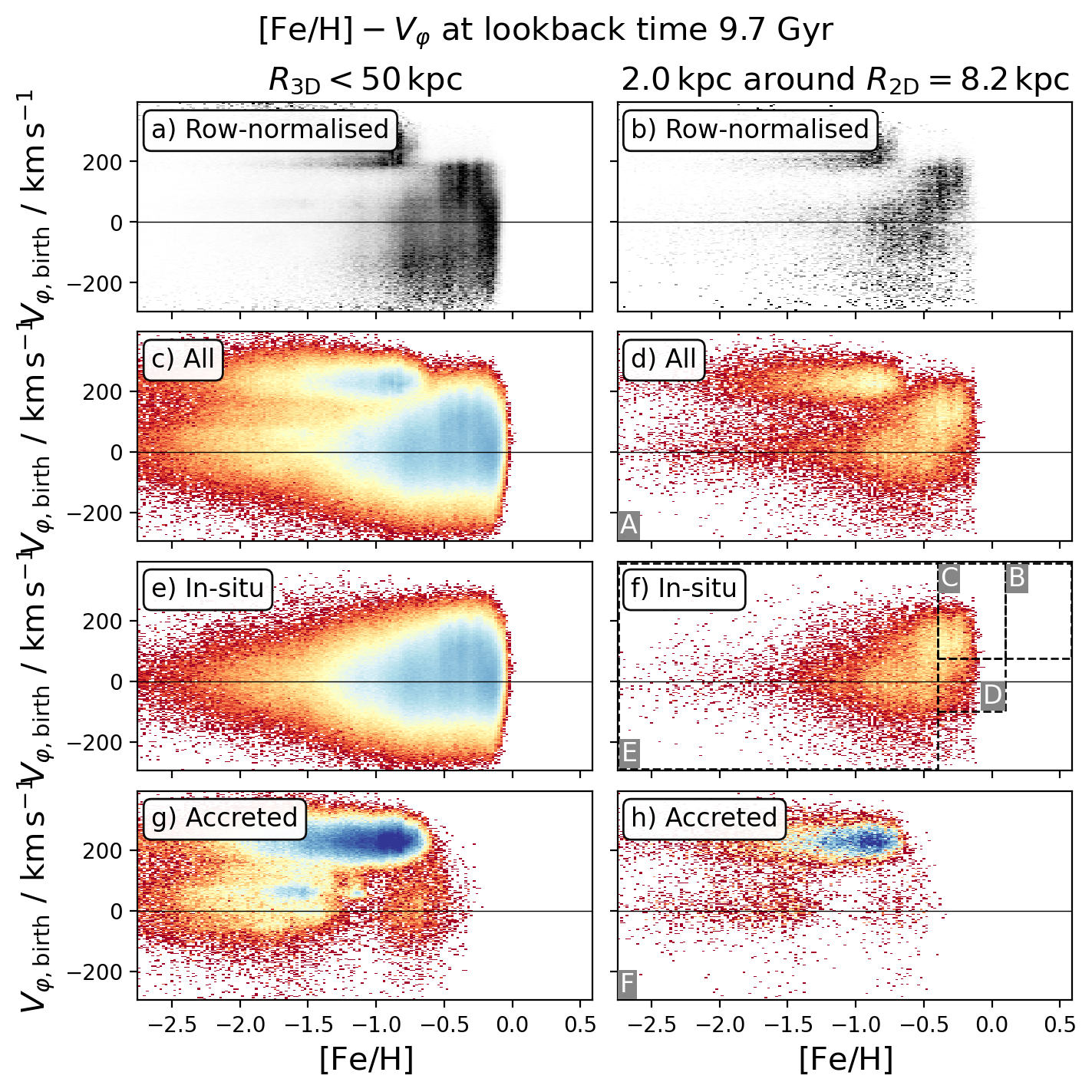}
\includegraphics[width=\columnwidth]{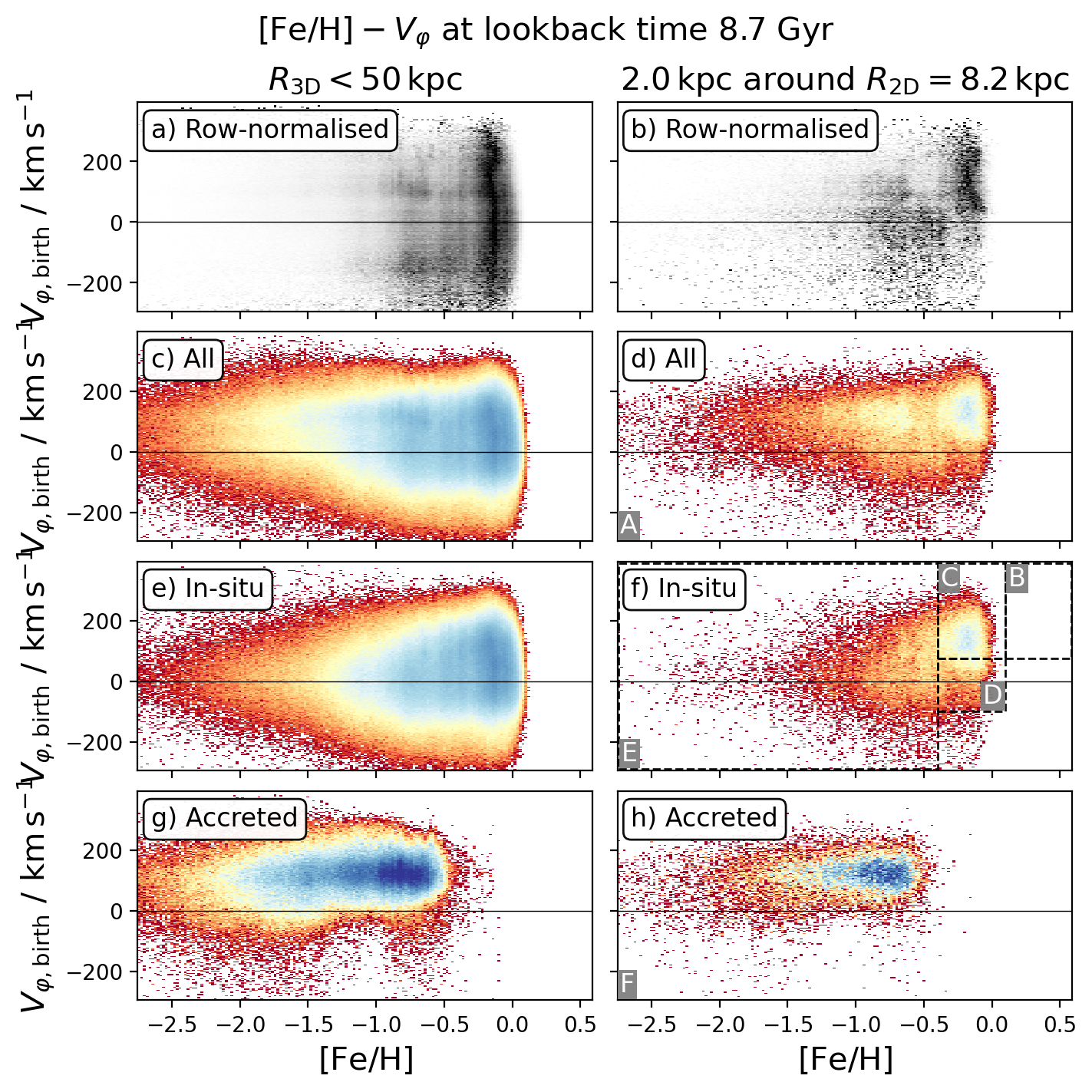}
\includegraphics[width=\columnwidth]{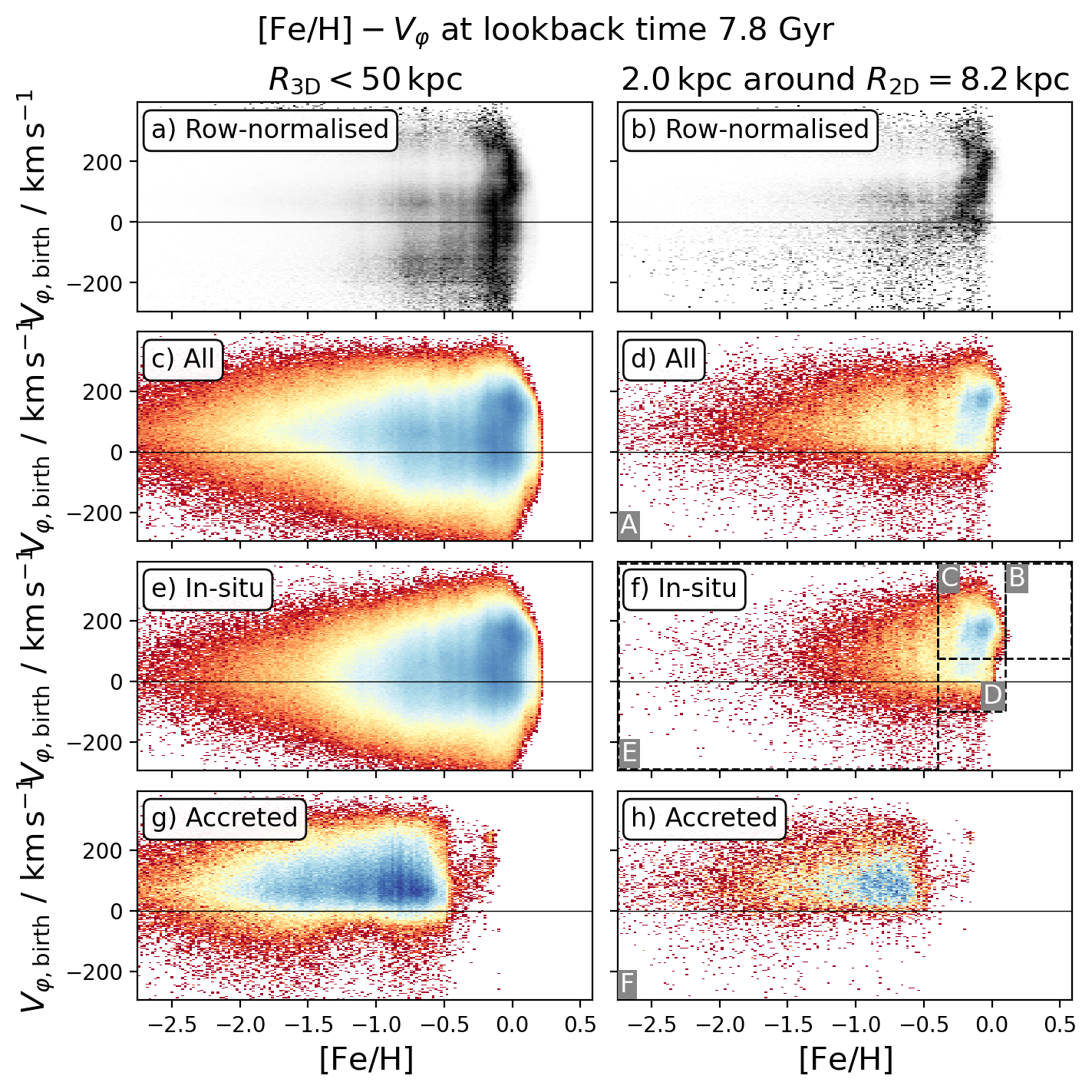}
\includegraphics[width=\columnwidth]{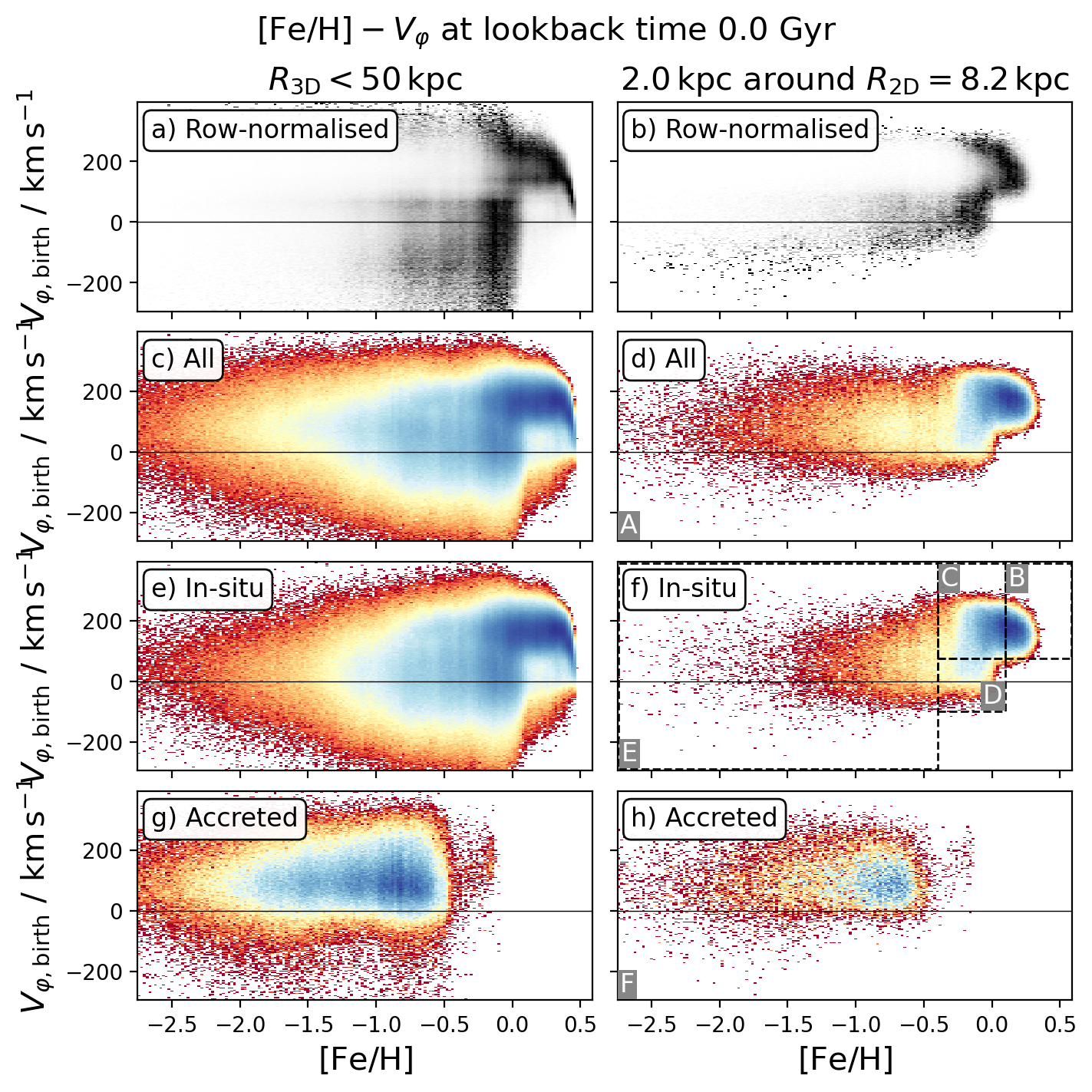}
\caption{
Evolution of the $\mathrm{[Fe/H]}$--$V_\varphi$ distribution of in-situ stars in the Solar-like neighbourhood at different stages of the merger interaction. 
Panels show the same diagnostic as Fig.~\ref{fig:splash_feh_vphi} for snapshots corresponding to lookback times of $9.7$, $8.7$, $7.8$, and $0\,\mathrm{Gyr}$.
}
\label{fig:splash_time_evolution}
\end{figure*}

To illustrate the state of the galaxy prior to and during the merger interaction, we reproduce the $V_\varphi$ and $\mathrm{[Fe/H]}$--$V_\varphi$ distributions shown in Figs.~\ref{fig:splash_vphi_distribution} and~\ref{fig:splash_feh_vphi} for several earlier snapshots of the simulation (Figs.~\ref{fig:splash_vphi_histograms} and~\ref{fig:splash_time_evolution}). The overall distribution remains broadly stable, while the Solar-neighbourhood subset shifts toward lower $V_\varphi$ after the merger. These snapshots demonstrate that a rotating but dynamically hot proto-disc is already present before the final coalescence of the merger.

The rotational support of the system increases steadily from $V/\sigma \approx 0.6$ at a lookback time of $\sim 9.7\,\mathrm{Gyr}$ to $V/\sigma \approx 1.1$--$1.3$ during the merger interaction, and continues to grow afterwards. This behaviour indicates that the merger interacts with an already existing proto-disc and contributes to the gradual build-up of rotational support.

To further quantify the impact of the merger on the Splash-like population, we examine the evolution of the $V_\varphi$ distribution for the same set of snapshots (Fig.~\ref{fig:splash_vphi_histograms}). For the full Splash-like population, the distribution remains broadly stable in terms of its mean and dispersion, with only minor changes in shape. In contrast, when selecting only those 4\% Splash-like stars that end up in the Solar-like neighbourhood at $z=0$, we observe a more pronounced evolution.

At a lookback time of $8.6\,\mathrm{Gyr}$, these stars follow an approximately Gaussian distribution with $V_\varphi = 120_{-70}^{+60}\,\mathrm{km\,s^{-1}}$. During the merger interaction, the distribution briefly shifts to higher values ($V_\varphi = 140_{-80}^{+70}\,\mathrm{km\,s^{-1}}$ at $8.5\,\mathrm{Gyr}$), likely reflecting the contribution of the accreted system with similar angular momentum. Subsequently, the distribution evolves toward lower angular momentum, reaching $V_\varphi = 80_{-70}^{+100}\,\mathrm{km\,s^{-1}}$ at $7.8\,\mathrm{Gyr}$ and $V_\varphi = 70_{-60}^{+70}\,\mathrm{km\,s^{-1}}$ at $z=0$.

This evolution indicates that the merger induces angular-momentum redistribution for a subset of stars, particularly those that populate the present-day Solar neighbourhood. At the same time, the overall stability of the full Splash-like population demonstrates that the merger does not uniformly reshape the entire distribution. Instead, the kinematic imprint of the merger appears to be spatially dependent, affecting only part of the population while leaving the bulk largely unchanged.

\label{lastpage}
\end{document}